\documentclass[useAMS,usenatbib]{mn2e}

\title[Fitting CMDs]{A maximum likelihood method for fitting colour-magnitude
diagrams}
\author[T. Naylor and R.D. Jeffries]{Tim Naylor$^{1}$ and
R.D. Jeffries$^{2}$\\
$^{1}$School of Physics, University of Exeter, Stocker Road, Exeter
EX4 4QL\\
$^{2}$Astrophysics Group, School of Physical and Geographical
Sciences, Keele University, Keele, Staffordshire, ST5 5BG}
\begin{document}

\date{}

\pagerange{\pageref{firstpage}--\pageref{lastpage}} \pubyear{2006}

\maketitle

\label{firstpage}

\begin{abstract}
We present a maximum likelihood method for fitting 
two-dimensional model distributions to stellar data in colour-magnitude space.
This allows one to include (for example) binary stars in an isochronal
population.
The method also allows one to derive formal uncertainties for fitted
parameters, and assess the likelihood that a good fit has been found.
We use the method to derive an age of $38.5^{+3.5}_{-6.5}$Myrs and a true
distance modulus of $7.79^{+0.11}_{-0.05}$ mags from the $V$ vs $V-I$ diagram of
NGC2547 (the uncertainties are 67 percent confidence limits, and the
parameters are insensitive to the assumed binary fraction).
These values are consistent with those previously
determined from low-mass isochronal fitting, and are the first
measurements to have statistically meaningful uncertainties.
The age is also consistent with the lithium depletion age of NGC2547, and the
HIPPARCOS distance to the cluster is consistent with our value.

The method appears to be quite general and could be
applied to any N-dimensional dataset, with uncertainties in each
dimension.
However, it is particularly useful when the data are sparse, in the
sense that both the typical uncertainties for a datapoint and the size of
structure in the function being fitted are small compared with
the typical distance between datapoints.
In this case binning the data will lose resolution, whilst the method
presented here preserves it. 

Software implementing the methods described in this paper is available
from http://www.astro.ex.ac.uk/people/timn/tau-squared/.

\end{abstract}


\section{INTRODUCTION}
\label{intro}

The extraction of astrophysical parameters from colour-magnitude
diagrams (CMDs), has been a crucial technique for astronomy 
since the discovery of the CMD as a diagnostic tool \citep[almost
certainly attributable to][]
{1911POPot..63.....H}.
Since a coeval population of singe stars occupies a curve in a CMD,
comparison with theoretical isochrones should allow a determination of
global properties of the population such as age, distance and
metallicity.
Unfortunately such determinations have been hampered by the lack of
good statistical methods for carrying out the comparison between
observation and theory.
For galactic astronomy, the main technique has been a visual
comparison of isochrones with the data (although more sophisticated
methods have been used for resolved populations in other galaxies).
This not only leads to questions of objectivity, but also makes it
impossible to derive statistically meaningful uncertainties for
parameter estimates.

Were the problem simply fitting a set of datapoints with
uncertainties in one dimension (say colour) to a curve then classical
$\chi^2$ analysis would suffice.
Unfortunately a datapoint in colour-magnitude space has uncertainties
in both colour and magnitude.
\citep[In addition the uncertainties are normally correlated, but 
as shown by][
this can
be overcome by transforming the problem into a magnitude-magnitude
space.]{1996ApJ...462..672T}
This problem can still be solved analytically if the curve is actually
a straight line \citep[][and references therein]{Nerit}.
\cite{1982ApJ...263..166F} extended this analytical approach to the
general case of a curve by a small curvature approximation.
Their method has been used on a significant volume of data, including
globular clusters \citep{1997A&AS..121..499B, 1993AJ....105.1420D} 
single-age extra galactic populations \citep{1999A&AS..134...21G} and
young ($<$ 10Myr-old) populations \citep{1997A&A...324..549T, 
1996A&A...312..499J}.
None of these studies make significant use of the uncertainty
measurements, partly
because of systematics, but partly there is also
the comment that they produce shallow $\chi^2$ spaces 
\citep{1986ApJ...307..738H} which result in large derived uncertainties
\citep{1991MNRAS.250..314N}.
This is clearly in part because the isochrones do not fit the data
well, but may also be a warning that, although one can place clusters
in an age sequence by eye, the absolute values of the ages, which must
be derived by comparison with the model isochrones, are not as precise
as we might hope.

There is a further limitation of the \cite{1982ApJ...263..166F}
technique, pointed out most explicitly by \cite{1998A&A...337..125G};
no population of stars consists entirely of single stars.
Unresolved binaries make up a significant fraction of most photometric
samples, and are seen as objects which lie up to 0.75 mags brighter
than the single star sequence.
Indeed, in some coeval populations a distinct equal-mass binary
sequence is observed 0.75 mags above the single-star sequence, with
unequal-mass binaries lying between the two.
Whilst one may be able to extract a single-star sequence by eye and
then fit it \citep{1992AJ....103..131H}, clearly the best way is to
fit the binaries as well.

Thus one arrives at the fundamental question we address in this paper.
If the model is a two dimensional distribution in the colour-magnitude
plane, can we derive a statistical test to fit the data to the model?
There has been some interest in using Bayesian methods to determine
the age of each star in a CMD \citep[][and references therein]
{2005A&A...436..127J}, and then using the mean for the cluster age.
Although \cite{2006astro.ph..3493V} demonstrate such a technique for
age determination, it is clear their work will be developed 
to fit other parameters as well.
The problem here, though, is the absence of a goodness-of-fit parameter to
choose between isochrones.
Another obvious solution is to bin the data into pixels, and compare this
with a model.
\cite{1997NewA....2..397D} and \cite{1997AJ....114..680A} have
developed this technique for large extra-galactic
populations, with \cite{2002MNRAS.332...91D} bringing much of the
literature together into a cohesive method.
The problem is, however, that our data are often sparse, by which we
mean the typical separation of datapoints is large compared with
their uncertainties (see Figure \ref{first_sim}).
Then binning the data simply has the effect of washing out our
hard-won photometric precision.

\cite{1996ApJ...462..672T} developed a technique which does retain the
datapoints as points, and which can been seen as a relative of the
method we use here.
They created simulated observations with a similar number of datapoints to
the observed dataset, and then used the distances in
colour-magnitude space between the points of the simulated and actual
observations as a fitting statistic.
Our method, first presented in \cite{2005prpl.conf.8502N}, improves on
this by using a quasi-continuous 2D distribution as the model, which
overcomes problems of sampling the model into a finite number of
datapoints, and allows robust uncertainties to be derived.

The method we are proposing is relatively intuitive, so rather than
embarking first on a formal analytical proof, we first give the
intuitive interpretation (Section \ref{intuitive}), and then discuss a
numerical experiment which demonstrates the technique using a simulated
observation, allowing us to conclude that it recovers the
correct answer and uncertainties (Section \ref{num_exp}).
The formal proofs are in Sections \ref{formal} and \ref{special}, and
the details of practical implementation in Section \ref{2d_approach}.
We draw all the work together in an example using real data in Section
\ref{2547}, before reaching our conclusions in Section
\ref{conclusions}.
An alternative to reading the paper in this order would be to gain a
working understanding from Sections \ref{intuitive}, \ref{num_exp},
and \ref{2547}, and try the worked examples available with the
software from\break 
http://www.astro.ex.ac.uk/people/timn/tau-squared.

\section{AN INTUITIVE INTERPRETATION}
\label{intuitive}

\begin{figure}
\vspace{70mm}
\includegraphics{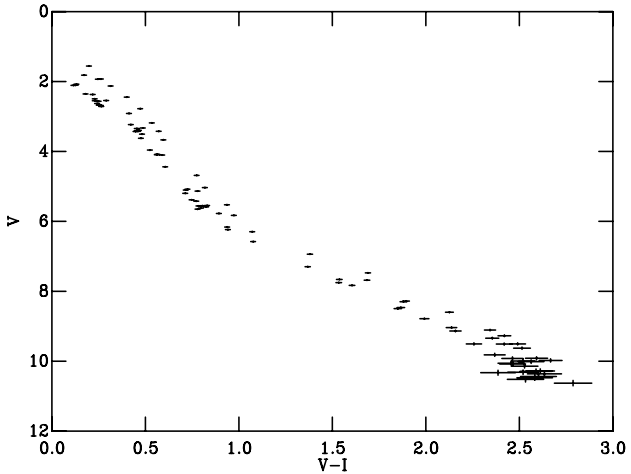}
\caption{
A simulated observation of a 40 Myr-old cluster.
See text for details.
}
\label{first_sim}
\end{figure}

\begin{figure}
\vspace{70mm}
\includegraphics{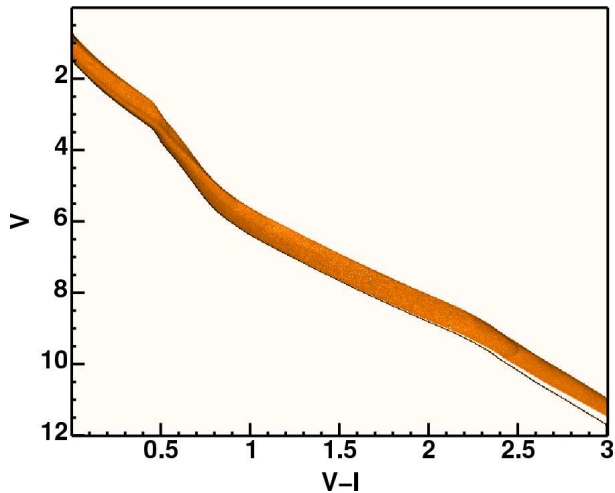}
\caption{
The expected distribution of datapoints underlying the simulation in
Figure \ref{first_sim}.
}
\label{dam97_40_fig}
\end{figure}

Figure \ref{first_sim} shows a simulated observation of 100 stars drawn from a 
40 Myr isochrone from \cite{1997MmSAI..68..807D}, henceforth referred to
as the DAM97 isochrones.
As for all the isochrones used in this paper we have converted the 
isochrones from effective temperature to colour-magnitude space using the
relationships derived from fitting the Pleiades \citep[see][for
details]{2001A&A...375..863J}.
The cluster is assumed to be unreddened and at a distance modulus of
zero.
The underlying power law mass function (d$N$/d$M \propto M^{-2.1}$) has been
chosen to give a reasonable spread of stars over the magnitude range
chosen.
We have assumed that 50 percent of the objects are unresolved binaries,
but that there are no higher order multiples.
Ignoring the higher order multiples should be a small effect since
only about 5 percent of systems have more than two members 
\citep{1991A&A...248..485D}. 
The masses of the secondary stars for the binaries are uniformly
distributed between the primary star mass and the lowest mass
available in the DAM97 models.
This is equivalent to assuming the mass-ratio distribution is flat.
Whilst there are many claims for structure in the distribution, after 
selection effects have been taken into account it is hard to argue
that a flat distribution is inconsistent with the data
\citep[e.g.][]{2003ApJ...599.1344M}.
In addition, as we shall show later the binary fraction, and by
implication mass ratio distribution, has little effect on the
parameters derived from the fits.
The presence of a low-mass cut-off in the DAM97 isochrones leads to
the empty wedge between the single star sequence and the
more equal-mass binaries visible below $V=9$ in Figure
\ref{dam97_40_fig}.
Stars in the wedge would represent binaries where the secondary star
lies below the lowest mass available in the models, and hence no stars 
can be placed in this region.
We shall show in Section \ref{extreme} that the wedge has a negligible
effect on our derived parameters.

Figure \ref{dam97_40_fig} shows the same model, but this time used to
generate many more objects, creating a surface density in
colour-magnitude space.
(For ease of display we have it renormalised such that the integral along each
horizontal row is one, but will ignore this renormalisation in what follows).
Were there no uncertainties in each datapoint, the relative
probability of there being a datapoint at some point $i$ at
$(c_i,m_i)$ is simply the value of Figure \ref{dam97_40_fig} at
$(c_i,m_i)$, which we refer to as $\rho(c_i,m_i)$.
Thus each datapoint has an associated value of $\rho$, and if we
multiply all these together, the resulting product, $D$ can be used as
a fitting statistic.
However, in analogy with $\chi^2$ we use $-2{\rm ln} D$, 
which we call $\tau^2$.
One can then consider moving the model around the plane in
colour and magnitude (or perhaps distance and reddening), until the value
of $D$ is maximised, or $\tau^2$ is minimised.

Introducing uncertainties for each datapoint does not have a large
impact on the method.
We introduce a two-dimensional uncertainty function for each
datapoint, which we call $U_i$ (for definiteness, one could consider
this to be a two-dimensional Gaussian).
One must now consider an uncertainty function centred at $(c_i,m_i)$,
and then integrate the product of $U$ and $\rho$ (the probability
distribution of Figure \ref{dam97_40_fig}), to obtain a probability
$P_i$.
We then calculate $\tau^2$ as $-2\prod{\rm ln}{P_i}$.
In fact, probably the most difficult problem is introduced by the
nature of the astronomical data; since the uncertainties in, say $V$ and
$V-I$ are correlated, we must actually integrate under two dimensional
Gaussians whose axis is skewed with respect to the colour-magnitude
grid (see Section \ref{correlation}).

It should be obvious from the above that this is a maximum likelihood
method.
As such it can be viewed as either Bayesian, or conventional
frequentist statistics.
As we discussed in the introduction it can be viewed as generalising
the method of \cite{1996ApJ...462..672T} to a model which provides a
continuous distribution.
As we shall show in Section \ref{into_chi}, it can also be viewed as a
generalisation of $\chi^2$. 

\begin{figure}
\vspace{70mm}
\includegraphics{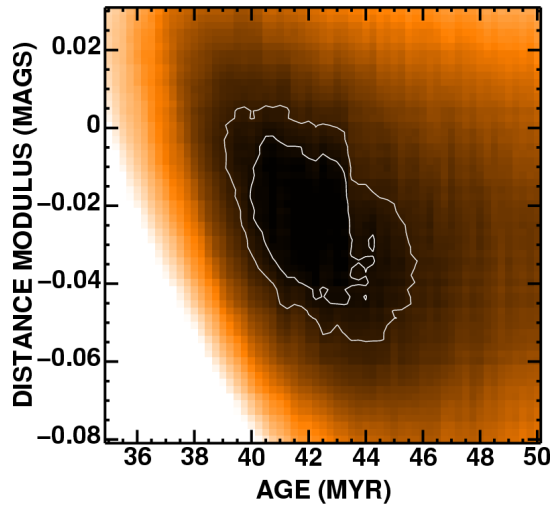}
\caption{
The $\tau^2$ space resulting from fitting the simulated observation in Figure
\ref{first_sim} (see Section \ref{uncer}).  
The values of $\tau^2$ are linearly scaled, and the white lines are
contours at the 67 percent ($\tau^2=317.3$) and 95 percent
($\tau^2=322.8$) confidence levels.
}
\label{first_sim_grid}
\end{figure}

\section{A NUMERICAL EXPERIMENT}
\label{num_exp}

Our numerical experiment was to find the age and distance
modulus of the artificial cluster described in Section
\ref{intuitive} from the simulated observation we described.
We followed the classical statistical path of finding the best fit to
the data, and hence derived estimated parameters (Section
\ref{fit_and_par}).
We then assessed whether this was a good fit (Section \ref{goodness}),
and then on the assumption it was, derived uncertainties in our fitted
parameters (Section \ref{uncer}).

\subsection{Fitting and parameter estimation}
\label{fit_and_par}

We compared our 100 datapoint simulated observation with a series of model
distributions with ages around 40Myr.
The model distributions we tested against used the same binary
fraction (50 percent) as the original simulation, and the same uniform
mass-ratio distribution.
We could have also used the same mass function as we used for the
simulation.
However, to do so would make this a highly unrealistic simulation of
fitting real data.
In practice, for deriving ages and distances the mass function is a
nuisance parameter. 
Whilst one may think that a simple power-law could be assumed over
the mass-range of interest, this would then have to be convolved with
the (often unknown) mass-dependent membership selection criteria.
For example, in Section \ref{2547} we shall use an X-ray selected
sample to determine the age of NGC2547, and the precise effect of X-ray
selection is unclear.
We therefore normalise our model distributions to have a constant number
of stars per unit magnitude (e.g. Figure \ref
{dam97_40_fig}).
We refer to this procedure as ``normalising-out'' the mass-function, 
and will discuss its implications in detail in Section
\ref{tau_prac}.
For datasets with well understood membership selection criteria our 
procedures can, in principle be simplified by removing the
normalising-out of the mass-function, allowing the mass function to be derived as well.

We tested several different offsets in magnitude for each age, which
yielded the $\tau^2$ space shown in Figure \ref{first_sim_grid}.
There is a minimum at 42.5Myr, a distance modulus of -0.0195 and
$\tau^2 = 311.9$, which is close to the values of 40Myr and 0.0 mags
of the artificial cluster from which the simulated observation was drawn.

\subsection{Goodness-of-fit}
\label{goodness}

\begin{figure}
\vspace{70mm}
\includegraphics{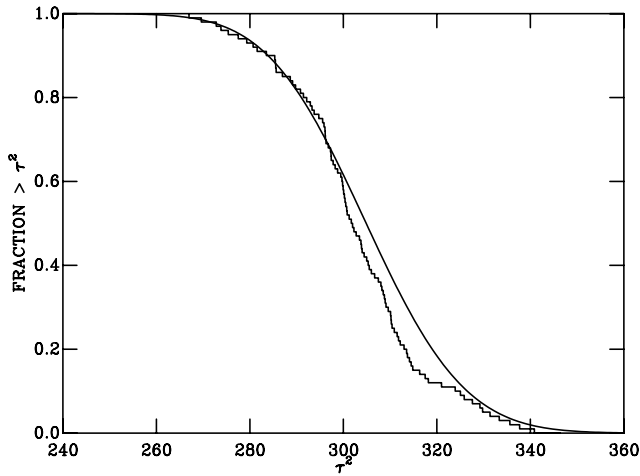}
\caption{
The smooth curve is the expected distribution of the probability of
obtaining a given $\tau^2$ calculated for the simulated observation (using
the best fit model) in the way described in Section \ref{tau}.
The histogram is the distribution of $\tau^2$ obtained by fitting 100
further simulated observations generated in the same was as the first. 
}
\label{sim_tau}
\end{figure}

In the case of $\chi^2$ fitting one uses the $F$-test, which in
essence is a prediction of the cumulative distribution of $\chi^2$.
We reject fits with $P_r(\chi^2)$ below some critical value, {\it
  e.g.} 5 percent.
(We use $P_r(x)$ throughout the paper to signify the probability
that a statistic exceeds the value $x$.
The subscript $r$ differentiates it from $P$, the
probability density in the colour-magnitude plane after
applying the uncertainty function.) 
It turns out that numerical integration allows us to predict, after
the fit is complete, the expected distribution of $\tau^2$ (Section
\ref{tau}), and thus assess the goodness of fit.
Such a distribution is the smooth curve in Figure \ref{sim_tau}.
For the example given we expect our value of $\tau^2=311.9$
to be exceeded in 34 percent of fits.
We can also normalise $\tau^2$ in a similar way to $\chi^2$ (for a
large number of degrees of freedom) by dividing
by the value we expect to be exceeded 50 percent of the time, which in
this case gives a reduced $\tau^2$, of $\tau_{\nu}^2=1.02$.

We can check that this is correct by creating a further 100
simulated observations, and examining the range of $\tau^2$ this
produces.
Figure \ref{sim_tau} shows (as a histogram) the distribution of
$\tau^2$ for from the 100 simulations.
The simulations suggest that 21 percent of observations would exceed 
$\tau^2=311.9$; clearly smaller than the 35 percent our theory
yields.
The reason for the difference is our normalising out of the mass-function
(see Section \ref{tau_prac}).
Despite this (which is a fundamental limit of the data, not of the
$\tau^2$ test), our method of calculating $\tau^2$ is
good enough to show that the fit is good, and of course
relative values remain useful for testing different models.

\subsection{Uncertainties for the parameters}
\label{uncer}

The simplest method of estimating the uncertainties would be to
create simulated datasets starting with the parameters derived from
the observation.
However, our normalising out of the mass-function precludes 
us from doing this.
We therefore produce bootstrap datasets by moving each datapoint at
constant magnitude onto the best-fit isochrone, and then adding to the
two magnitudes a random offset drawn from a population with the
appropriate Gaussian distribution for the error bars associated with
the datapoint.
Since there is not a unique isochronal colour associated with each 
magnitude (because of the effects of binarism), we have to assign the
datapoint to a position in colour using the relative likelihood of 
each colour drawn from the model.
Hence we have assumed that the probability of any given combination of
parameters being the correct one is identical to the probability of
obtaining those parameters if the underlying model was actually the
best-fitting model.  
We then make 100 bootstrap datasets, and examine the resulting
values of the derived parameters, using the RMS about the mean value
as the uncertainty.
This gives uncertainties in distance modulus and age of 0.012 mags
and 1.1 Myr respectively.

We can test these estimates of the uncertainties using
the 100 simulated observations we created for
Section \ref{goodness}.\footnote{  
  The situation becomes unavoidably confusing at his point, as we now
  have two simulated groups of observations, each of 100 realisations.
  We refer to the 100 simulated observations created for Section
  \ref{goodness} in the same way as our original simulated
  observation, as ``simulated observations''.
  The 100 simulated datasets created in Section \ref{uncer}, which in
  a real case would be derived from the observation by forcing all the
  data back onto the isochrone and then scattering them according to
  their uncertainties we refer to as ``bootstrap datasets''.}
These give a scatter in distance modulus
and age of 0.011 mags and 0.9 Myr, in good agreement with our
bootstrap method for determining the uncertainties.

In practice, we are interested in more than the simple uncertainties, as
there is a correlation between distance modulus and age.
We deal with this in an analogous way to $\chi^2$ fitting by drawing a
contour in the $\tau^2$ space which encloses a given fraction of the
probability of where the solution lies.
We take the values of the distance modulus and age derived from
each bootstrap dataset, and find the corresponding value of
$\tau^2$ in our $\tau^2$ space derived from the
first simulated observation (Figure \ref{first_sim_grid}).
(Not the value of $\tau^2$ given by the fit to the bootstrap
datasets.)
This allows us to draw the contour at constant $\tau^2$ (317.3) which
encloses 67 percent of the derived values, i.e. a ``1$\sigma$''
confidence limit.
This is plotted in Figure \ref{first_sim_grid} and shows the expected
correlation between age and distance modulus.
\footnote{Note that the $\tau^2$ for the 67 percent confidence contour 
is not given by $1-P_r(\tau^2)=0.67$.  
The analogous case for $\chi^2$ is that for one free parameter one
uses the minimum $\chi^2$ + 1 as the confidence contour.}

Again we can test this using our simulated observations from Section
\ref{goodness}.
We take the values of the distance modulus and age derived from
each simulation, and find the values associated with them from
the $\tau^2$ space derived from the first simulated observation.
67 percent of them lie below $\tau^2$=317.9.
Given that we have 100 simulations, we actually require the $\tau^2$
below which 67 percent of some large parent population lies, which we
estimate is between 317.2 and 318.1.
This range which includes the value derived using our proposed technique,
implying that technique is correct.

We also tried using a more traditional bootstrap method
\cite[e.g.][]{2003psa..book.....W}. 
For such a bootstrap to work the values of the datapoints (or in our
case the values calculated from them) must be identically distributed
\citep[see for example Section 15.6 of][]{1992nrfa.book.....P}.
It is quite clear that the ages and distance moduli derived from each
datapoint are not identically distributed, but the traditional
bootstrap often works sufficiently accurately even when this
assumption is quite strongly violated.
To see if this was the case, we created 100 new data sets by randomly
selecting 100 points from the original data.
(Thus, as is normal in such a bootstrap method, a significant fraction
of the datapoints in each realisation are the same.)
We found the RMS for each parameter using this dataset, which yields
uncertainties of 1.2 Myr and 0.011 mags.
Again, these are consistent with those calculated using our method.
However, we found that the suggested 67 percent confidence contour 
for $\tau^2$ is too low (314.7).
To check that this failure of the traditional bootstrap was not due
to our normalising out the mass function we performed a similar
simulation using models which retained the mass function.
Again we found our bootstrap gave a similar confidence interval to
that implied by many simulated observations of the same ``cluster'',
but in this case the traditional bootstrap overestimated the
uncertainties.
Clearly the derived parameters from each datapoint are not
sufficiently close to identically distributed for the traditional
bootstrap to work.

\subsection{Conclusion}

In this section we have validated the $\tau^2$ test by simulating a
dataset and recovering the original parameters.
We have also shown that we can estimate reliable uncertainties in
the measured parameters by creating bootstrap datasets.
The ``base'' for the simulations is created by moving each point in 
colour space until it lies on the isochrone.
By examining the range of $\tau^2$ the values of the parameters
derived from these datasets, we can estimate confidence intervals
analogous to those used in $\chi^2$ analysis.

\section{FORMAL DEFINITION}
\label{formal}

\begin{figure}
\vspace{70mm}
\includegraphics{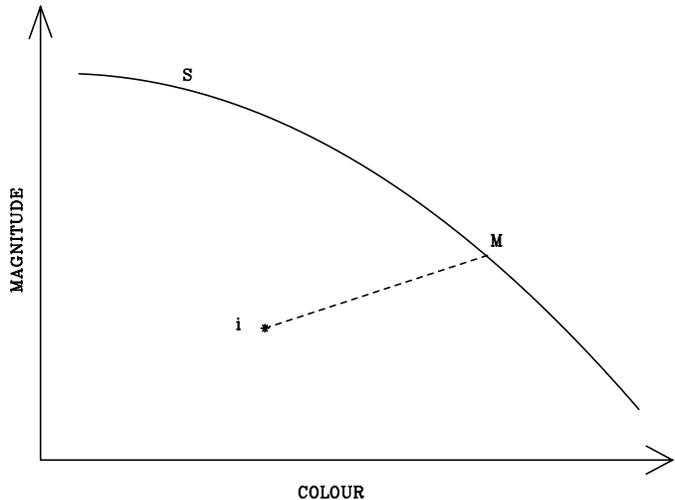}
\caption{
A schematic showing a sequence $S$, an observed datapoint $i$ and a
point $M$ which may be its position unperturbed by observational error.
}
\label{fj}
\end{figure}

Having shown by numerical experiment that $\tau^2$ can work, we must
now put it on a formal mathematical footing.
Figure \ref{fj} shows a colour magnitude plane, with a sequence $S$
and an observed datapoint $i$ at $(c_i,m_i)$.
The datapoint will have an associated two-dimensional
probability distribution, which we will assume is Gaussian.
This allows us to calculate the probability that the true values of
$c$ and $m$ lie within any specified range.
Thus if the datapoint lies at ($c_i, m_i$), the probability that the
true value lies within an elemental box of area ${\rm}dc\ {\rm d}m$ about $M$ at ($c,m$) can
be written as $U_i(c-c_i, m_i-m){\rm d}c\ {\rm d}m$, where $U$ is a 2D function
which represents the uncertainty for a given datapoint.

We now assume that we have a model $\rho$ which predicts the true
density of stars in the colour-magnitude plane.
If that model is, say, a $\delta$-function at ($c,m$), then the
probability that our data originates from the model is simply the
integral of the product of the $\delta$-function and $U_i$.
More generally, the likelihood for any given datapoint $i$ is given by
\begin{equation}
P_i=\int{U_i(c-c_i, m-m_i)\rho(c,m){\rm d}c\ {\rm d}m}.
\label{hypothesis}
\end{equation}
If there are $N$ datapoints, the likelihood that the whole
distribution originates from the model is the product of the
probabilities for each point.
\begin{equation}
D=\prod_{i=1,N}{P_i}=\prod_{i=1,N}\int{U_i(c-c_i, m-m_i)\rho(c,m){\rm
    d}c\ {\rm d}m}
\label{for_D}
\end{equation}
If we now define $\tau^2$ as -2ln$D$, then we arrive at the formal
definition of $\tau^2$,
\begin{equation}
\tau^2=-2\sum_{i=1,N}{\rm ln}\int{U_i(c-c_i, m-m_i)\rho(c,m){\rm d}c\ {\rm d}m}.
\label{for_tau}
\end{equation}
For most practical applications $U_i$ has Gaussian uncertainties and
is given by
\begin{equation}
U_i(c-c_i, m-m_i)=e^{-{(c-c_i)^2\over{2\sigma_{c_i}^2}}
                     -{(m-m_i)^2\over{2\sigma_{m_i}^2}}},
\label{gauss_uncer}
\end{equation}
where $\sigma_{c_i}$ and $\sigma_{m_i}$ are the uncertainties in each
measurement.

There are two obvious interpretations of equation \ref{for_tau}.
The first is that one has simply taken the model and blurred it by the
uncertainties in each datapoint.
The likelihood is then simply the product of the values of the model
at each point.
Alternatively, we have integrated model probability under the 2D 
Gaussian uncertainty surface.
In either interpretation the process of maximising this function to
obtain the best fit is analogous to maximising the cross correlation
function, though one uses the product rather than the sum of the
individual pixel values.

\section{SPECIAL CASES}
\label{special}

Before using our full two dimensional implementation of $\tau^2$ it is
useful to reduce Equation \ref{for_tau} for three special cases.
These show how (i) $\tau^2$ is related to $\chi^2$;
(ii) that it gives the standard form for fitting a straight line to data with
uncertainties in two dimensions; and (iii) 
that it can also reduce to the
same approximation as that of \cite{1982ApJ...263..166F} for curve fitting
with uncertainties in two dimensions.

\subsection{Curve fitting for data with one dimensional uncertainties}
\label{into_chi}

The most important special case to derive is that for when the model
predicts that a point whose true value is ($c,m$) should always have an
observed value of $c_i=c$, but has a range of possible
observed values $m_i$, represented by a Gaussian probability distribution.
In this case $\tau^2$ should behave like $\chi^2$.
Removing the dependence on $c$ from Equations \ref{for_tau} and 
\ref{gauss_uncer} yields
\begin{equation}
\tau^2=-2\sum_{i=1,N}{\rm ln}\int{e^{-{(m-m_i)^2\over{2\sigma_{m_i}^2}}} \rho(m){\rm d} m}.
\end{equation}
Further, for any single datapoint $\rho(m)$ is a $\delta$ function
centred on $m$, and with a normalisation we choose to be one, thus
\begin{equation}
\tau^2=-2\sum_{i-1,N}{\rm
  ln}\Big(e^{-{(m-m_i)^2\over{2\sigma_{m_i}^2}}}\Big)
=\sum_{i-1,N}{(m-m_i)^2\over{\sigma_{m_i}^2}}.
\end{equation}
This is the standard form for $\chi^2$ fitting to a function with
uncertainties in one dimension.
Thus we have shown that $\chi^2$ is a special
case of $\tau^2$, where the model is a curve and the data have
uncertainties in one dimension.

\subsection{A linear isochrone}
\label{linear}

We now wish to examine the special case where the probability
distribution is a linear sequence, but the data now have uncertainties
in both co-ordinates.
We have three aims in presenting this special case.
First, to show that the standard form for fitting a straight line with 
uncertainties in two dimensions is a special case of $\tau^2$.
Second, we will
test our (numerical) implementation of $\tau^2$ by
checking it recovers the same answer as the analytical expression.
We find that if this is to be the case we must use the correct
normalisation for $\rho$, which we derive below.
Finally, there is an intuitive interpretation of the analytical
expression which is useful for interpreting the more general case of
fitting a curve with uncertainties in both dimensions.

\subsubsection{Analytical form}

Formally we wish to assess the probability that a point $i$ at ($c_i,m_i$)
originates from the isochrone
\begin{equation}
m = {{{\rm d}m}\over{{\rm d}c}}c + k,
\label{lin_iso}
\end{equation}
where $k$ is a numerical constant.
We begin by changing to a co-ordinate system ($x,y$), a process shown
graphically in Figure \ref{sl}.
We first normalise by the uncertainties in each axis, then place
($c_i,m_i$) at the origin, and finally rotate the system through 
an angle $\theta_i$ such that the x-axis lies parallel to the sequence.
(We use the subscript $i$ to emphasize that $\theta$ depends on the
uncertainties and so is potentially different for each datapoint.)
Thus
\begin{eqnarray}
{{m-m_i}\over{\sigma_{m_i}}} & = & y{\rm cos} \theta_i + x{\rm
  sin}\theta_i,
\label{trans1}\\
{{c-c_i}\over{\sigma_{c_i}}} & = & x{\rm cos} \theta_i - y{\rm
  sin}\theta_i. 
\label{trans2}
\end{eqnarray}
Equation \ref{hypothesis} then becomes
\begin{equation}
P_i=\int{e^{-{{x^2+y^2}\over2}}}\rho(x,y){\rm d} x {\rm d} y.
\end{equation}
In this co-ordinate system we denote the $y$-distance between the
$x$-axis and the sequence $y_0$.
We can now divide $\rho(x,y)$ into $\rho(x)\rho(y)$ where $\rho(x)$ is
constant and $\rho(y)=0$ except where $y=y_0$.
This allows us to separate the integrals, and find that
\begin{eqnarray}
P_i&=&\int{e^{-{{y_0^2}\over2}}e^{-{{x^2}\over2}}\rho(x)\rho(y){\rm d} x {\rm d} y}\\ 
   &=&e^{-{{y_0^2}\over2}}\rho(x)\int{e^{-{{x^2}\over2}}{\rm d}x}\int{\rho(y){\rm d} y} \\ 
   &=&\sqrt{2\pi}e^{-{{y_0^2}\over2}}\rho(x)\int{\rho(y){\rm d} y}.
   \label{lin_iso_prob}
\end{eqnarray}
Now $\rho(x)\int{\rho(y){\rm d} y}$ is
the number of objects per unit length in $x$, and in
terms of the number of objects per unit magnitude, $\rho(m)$, is
$\sigma_{m_i}$sin$\theta_i \rho(m)$, thus
\begin{equation}
P_i = \sqrt{2\pi}e^{-{{y_0^2}\over2}} \sigma_{m_i} {\rm sin}\theta_i \rho(m).
\end{equation}
Thus, Equation \ref{for_tau} becomes
\begin{equation}
\tau^2 = \sum_{i=1,N}{y_0^2} - 2\sum_{i=1,N}{{\rm ln}(\sigma_{m_i} {\rm sin
 }\theta_i \rho(m)\sqrt{2\pi}}).
\label{linear_final}
\end{equation}

\subsubsection{Intuitive interpretation}
\label{intuitive_int}

This equation has an intuitive interpretation, which is especially
useful for what follows.
The probability that a star at ($c_i, m_i$) originates from a given point
on an isochrone is given by the probability that there is a datapoint
whose true value lies at that point on the isochrone, multiplied by
the probability that the uncertainties could move it to ($c_i, m_i$).
For the whole isochrone, therefore, the probability that it will yield
a point at ($c_i, m_i$) is given by the
line integral along the isochrone, multiplied at each point by the
probability of it being moved to ($c_i, m_i$). 
This probability distribution is (in normalised units) simply a
two-dimensional Gaussian distribution centred on ($c_i, m_i$). 
Any linear cut through such a 2D Gaussian, such as that
made by the isochrone, is itself a 
1D Gaussian, but with is peak reduced by $e^{-{y_0^2\over2}}$ with
respect to the 2D distribution.
Thus the integral along the line is the integral under this 1D
Gaussian.
The ratio of the integrals under 1D and 2D Gaussians of equal peak
height is $\sqrt{2\pi}$, but this must also be multiplied by the
decrease in peak, $e^{-{y_0^2\over2}}$, explaining the form of Equation 
\ref{lin_iso_prob}.

\subsubsection{Testing the linear isochrone}
\label{linear_test}

\begin{figure}
\vspace{70mm}
\includegraphics{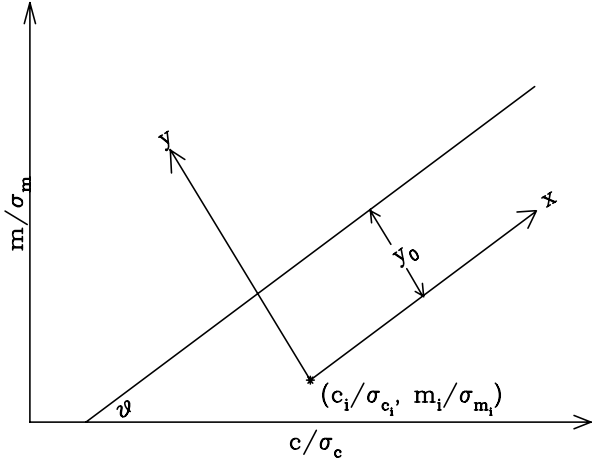}
\caption{A linear isochrone which makes an angle $\theta$ with axes
normalised by the uncertainties in each dimension.
The $x$ and $y$-axes define a rotated co-ordinate system parallel
with the isochrone, centred on the datapoint ($c_i$, $m_i$). 
Note magnitude axis is reversed.}
\label{sl}
\end{figure}

Equation \ref{linear_final} gives us a practicable way of fitting a
linear isochrone, by minimising $\tau^2$ as a function of $y_0$ and
the gradient of the isochrone (which is related to $\theta_i$).
First, if we wish $\tau^2$ to reduce to $\chi^2$ we must choose the
normalisation of $\rho$ in Equation \ref{linear_final}.
Since $y_0$ is distributed as a Gaussian with a standard deviation of one, this
means we must ensure the second term is zero.
Thus
\begin{equation}
\sum_{i=1,N}{{\rm ln}(\sigma_{m_i} {\rm sin}\theta_i \sqrt{2\pi}}) + 
\sum_{i=1,N}{\rm ln}(\rho(m))=0,
\end{equation}
giving
\begin{equation}
\rho(m)^{-N} = \prod_{i=1,N}(\sigma_{m_i} {\rm sin }\theta_i \sqrt{2\pi}).
\label{rho_norm}
\end{equation}
From Figure \ref{sl} it is clear that $\theta$ is related to the
gradient of the isochrone by
\begin{equation}
{{{\rm d}m}\over{{\rm d}c}}={{\sigma_{m_i}}\over{\sigma_{c_i}}}{\rm
  tan}\theta_i.
\end{equation}
The value of $y_0$ can be found using the above equation, and
setting
$x$=0 in Equations \ref{trans1} and \ref{trans2}, and
substituting into Equation \ref{lin_iso} to obtain
\begin{equation}
y_0 = 
{c_i\over\sigma_{c_i}}{\rm sin}\theta_i +
  {{k-m_i}\over{\sigma_{m_i}}}{\rm cos} {\theta_i}.
\label{y_0}
\end{equation}
For a given linear isochrone and set of simulated datapoints this
means we can calculate analytically a value for $\tau^2$.
We can then use this to test the 2D numerical code we describe below.

\subsubsection{Comparison with a straight line fit with 2D uncertainties}

Clearly the best-fitting straight line will be obtained by minimising
the sum over all datapoints of $y_0^2$ in Equation \ref{y_0}.
We can rewrite the equation such that
\begin{equation}
y_0^2={{{c_i {{{\rm d}m}\over{{\rm d}c}} + k - m_i}^2}\over
      {\sigma_{m_i} ^2 + \sigma_{c_i} ^2 {{{\rm d}m}\over{{\rm d}c}}}}.
\end{equation}
This is the standard expression to be minimised for fitting a straight
line to data with uncertainties in both co-ordinates
\citep[e.g.][]{Nerit}, which demonstrates that such fitting is a
special case of $\tau^2$.

\subsection{A real isochrone}
\label{real}

We can use the interpretation of Equation \ref{linear_final} presented
in Section \ref{intuitive_int} to 
visualize the limit in which the approximation that
the isochrone is linear is no longer valid.
Once the curvature of the isochrone becomes large compared with the
typical uncertainties for a datapoint, then it cannot be approximated
to a straight line when the line integral is performed.
However, for the case where the curvature is small, one might still be
able to use Equation \ref{linear_final}, interpreting $y_0$ as
the distance of closest approach of the line to the datapoint, and
$\theta_i$ referring to the gradient of the isochrone at closest approach.
Although a rather different derivation, such a technique would be
identical, save some normalisation factors, to the near-point
estimator of \cite{1982ApJ...263..166F}.

\section{THE TWO DIMENSIONAL APPROACH}
\label{2d_approach}

\subsection{Implementation}
\label{imp}

We can gain our first insights into the 2D case by reproducing the
results from the 1D-linear and 1D-real isochrones of Sections
\ref{linear} and \ref{real} using the 2D algorithm.

We evaluate the integral in Equation \ref{for_tau} using a 2D
grid.
We represent $\rho(c,m)$ as a grid and, for reasons we will discuss
later, populate this grid by a Monte Carlo method.
For these 1D isochrones we begin by randomly selecting a magnitude, and
then assigning a colour according to the isochrone.  
The value of the appropriate pixel of $\rho(c,m)$ is then incremented by one.
At the end of the Monte Carlo we then ensure that $\rho(m)$ is one by
dividing each pixel by the sum of all pixels at that magnitude.   
This means that in practice the initial distribution in magnitude used
by the Monte Carlo is unimportant, provided it is smooth.

For each datapoint we can now evaluate Equation \ref{hypothesis}.
We multiply each pixel of $\rho(c,m)$ by Equation \ref{gauss_uncer}.
In principle, before using $\rho$ we should correct it by the
normalisation given in Equation \ref{rho_norm}.
In practice it is simpler to apply a correction to the 
$\rho$ used for each datapoint, which when the probabilities for each
datapoint are multiplied together gives the same effect.
Thus for each datapoint we divide $\rho$ by the normalisation factor 
$ \sigma_{m_i} {\rm sin }\theta_i \sqrt{2\pi}$.
To evaluate ${\rm sin }\theta_i$ we require the gradient at each
pixel, which we evaluate and store at the same time as we calculate
$\rho(c,m)$, by differencing the $c$ and $m$ values of the most
extreme valued points from the Monte Carlo which lie within the pixel.
Of course the gradient is only defined on the isochrone, and we need
it for a general point in the CMD.
We can be arbitrary about how we make this generalisation, since our
normalisation is only there to ensure that if we have a straight line
(where the gradient is obviously always the same)
we obtain a $\tau^2$ of one per datapoint.
We therefore choose the gradient for an arbitrary pixel to be that of the 
isochrone at the magnitude of the pixel.
At this point one can test the code is functioning correctly by using
a linear isochrone and testing the result for $\tau^2$ against the 
analytical result given in Equation \ref{linear_final}.

To move to the more general 2D case one fills the array for $\rho$
by selecting stars randomly according to some mass function. 
They are assigned companions (or not) according to the preferred binary 
frequency and mass functions, and then one uses isochrones to place the
resulting systems in colour-magnitude space.
The remainder of the procedure is as before for the linear isochrone.

\subsection{Correlated Uncertainties}
\label{correlation}

A significant issue with any CMD is that the
uncertainties are correlated, since a change in, say $V$ also results
in a change in $V-I$.
Perhaps the most obvious change in formalism to deal with this is that
suggested by \cite{1996ApJ...462..672T}, where the actual fitting is
carried out in a magnitude-magnitude space.
We have found it simpler (and therefore more robust against coding
errors) to use colour-magnitude space throughout our code.
However, at the point of evaluating Equation \ref{gauss_uncer} one can
calculate the argument of the exponential in magnitude-magnitude
space, reconstructing the uncertainty in the second magnitude using
the uncertainties in magnitude and colour.
In principle this allows considerable flexibility, including the
ability to deal correctly with data which have been created using a
colour term in the transformation from instrumental
to apparent magnitude, and a co-efficient other than unity in the
transformation from instrumental to apparent colour.

\subsection{The distributions of $\tau^2$}
\label{binaries}

Once we have fitted our data, to calculate whether it is a good fit we
need to know $P_r(\tau^2)$.
To calculate this we must first calculate the distribution for a
single point, and then calculate the expected distribution for the
whole ensemble of datapoints. 

\subsubsection{The $\tau^2$ distribution for one datapoint}
\label{tau}

To understand the form of the $\tau^2$ distribution it is useful to
begin by considering a classical $\chi^2$ fitting problem, but
solved as though it were suitable for $\tau^2$.
In such a problem the model isochrone is a curve in colour magnitude
space, and the uncertainties are treated as 2D Gaussians which are
infinitely thin in the colour dimension, and have the correct width in
magnitude space to represent the 1D uncertainty.
We can calculate the chance that a star at a given point on the
sequence actually appears, due to observational error, at a given 
position on the CMD.
If we integrate this along the entire sequence we obtain the
probability of there being a datapoint at any given position on the
CMD.
Since our uncertainties are Gaussian, and the line is a form of
$\delta$-function, the 
distribution of probabilities in the plane is itself Gaussian.
Thus, the likelihood of finding a datapoint at given probability, say
$P$, is proportional to the fraction of the CMD covered by pixels with
that probability, multiplied by $P$.
Assuming all pixels have the same area, this can be calculated
numerically by summing the values of all pixels for which $P$ lies
within a given (infinitesimal) range.
Strictly speaking this should only be interpreted in the cumulative
sense, i.e. that the probability of finding a datapoint with a
probability of $P$ or less is proportional to the fraction of the area 
covered by each probability less than $P$, multiplied by that
probability, and then integrated over all probabilities less than $P$.
We, of course, have chosen not to work in probability $P$, but
in $\chi^2 = -2 {\rm ln} P$.
Thus we have not quoted the chance of a datapoint being at a probability
$P$ or less, rather the chance of it lying at a given $\chi^2$ or
more.

\begin{figure}
\vspace{70mm}
\includegraphics{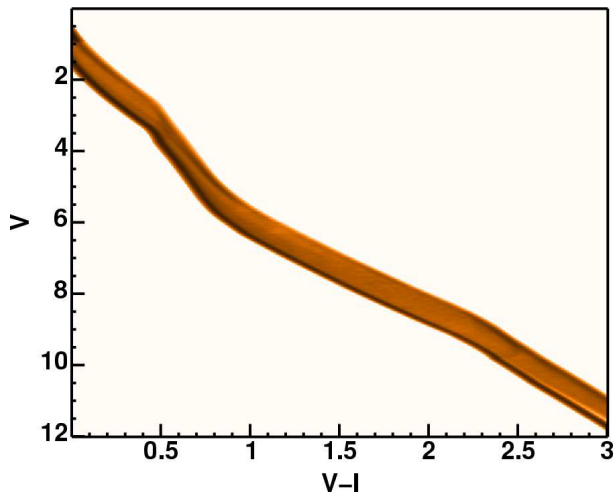}
\caption{
A model created using the DAM97 isochrones, and a binary fraction of
0.5.
The uncertainties used are 0.01 mag in each filter.
}
\label{dam97_0.01}
\end{figure}

Of course when we perform this sum over the plane, the resulting
distribution will be that of $\chi^2$.
We can still retain a $\chi^2$ distribution in the plane if we make
the uncertainties two dimensional, provided we restrict the isochrone
to be a straight line.
But if the model is to be a curve, and/or include binaries, the
distribution of probability in the plane will no longer be Gaussian,
and the probability of exceeding a given value will no longer behave
like $\chi^2$.
We can still accurately predict the distribution of values of 
$-2 {\rm ln} P$ we expect to get.
That is obtained by simply creating a histogram of the probabilities
from an image such as that in Figure \ref{dam97_0.01}.
But, this will no longer be distributed as $\chi^2$, and to emphasize
this fact we will now refer to $ -2 {\rm ln} P$ as $\tau^2$.

\begin{figure}
\vspace{70mm}
\includegraphics{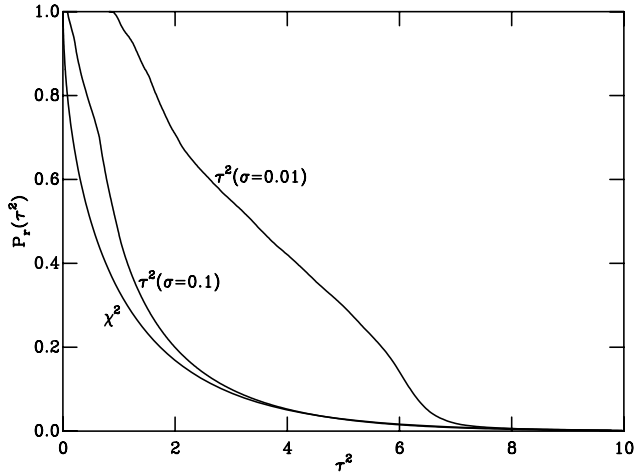}
\caption{
The $\tau^2$ distribution (the probability of exceeding a given
$\tau^2$ from the DAM97 isochrones using two different
values of the uncertainty.
A $\chi^2$ distribution for one degree of freedom is shown for comparison.
}
\label{tau_integ}
\end{figure}

\begin{figure}
\vspace{70mm}
\includegraphics{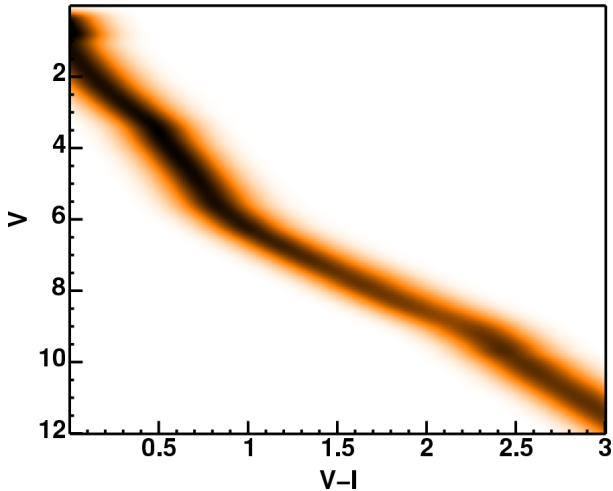}
\caption{
A model created using the DAM97 isochrones, and a binary fraction of
0.5.
The uncertainties used are 0.1 mag in each filter.
The feature at bright magnitudes is caused by the upper cut-off in
mass for the DAM isochrones.  
At this cut-off the binary sequence
rises to brighter magnitudes than the single star sequence.
When the single-star sequence ends, the smoothed sequence moves
redwards.
}
\label{dam97_0.1}
\end{figure}

In Figure \ref{tau_integ} we show the cumulative distribution for the
value of $\tau^2$ taken from Figure \ref{dam97_0.01}, where the
uncertainties are 0.01 magnitudes in each filter.
When compared with the $\chi^2$ distribution for one degree of
freedom, there are two major differences between $\tau^2(\sigma=0.01)$
and $\chi^2$.
First $\tau^2(\sigma=0.01)$ has no values below about 1, and second it
falls much more slowly.
The slow fall is the effect of the ``plateau'' region between the single and
binary star sequences, which contributes a large area of low
probability, and hence high $\tau^2$.
The absence of values below about one is the result of our requirement
that $\rho(m,c)$ at a given magnitude integrates to one over all colours.
This imposes a maximum value on $P$, and hence a minimum on $\tau^2$.
As one moves to larger uncertainties ($\tau^2 (\sigma = 0.1)$ in
Figure \ref{tau_integ}), these differences become less pronounced.
The change from $\tau^2 (\sigma = 0.01)$ to $\tau^2 (\sigma = 0.1)$
shows that as the uncertainties become larger, $\tau^2$ tends to
$\chi^2$.
The reason for this is clear if one compares Figure \ref{dam97_0.01}
with Figure \ref{dam97_0.1}.
As the uncertainties become large 
compared with the distance between the single and binary star
sequences, we can approximate them to a single sequence. 

\subsubsection{The $\tau^2$ distribution for many datapoints}

Having calculated $\tau^2$ for a single datapoint, it would appear
straightforward to calculate it for an ensemble.
We will do this by comparison with the case for $\chi^2$.

\begin{figure}
\vspace{70mm}
\includegraphics{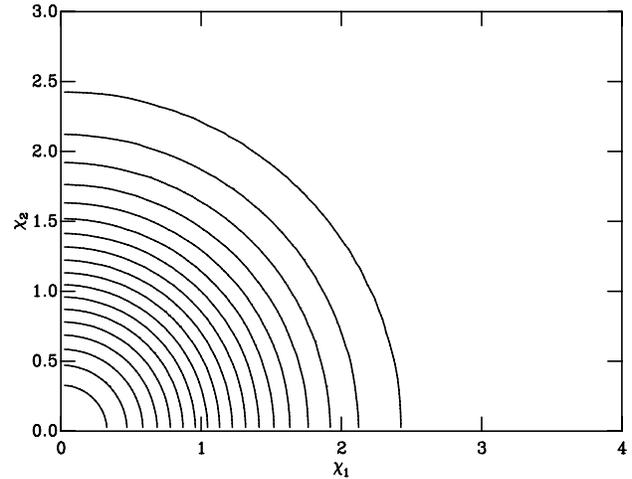}
\caption{
The two-dimensional differential probability distribution of $\chi$.
The contours are evenly spaced starting at a probability of zero.
Note that the axes are in $\chi$ not $\chi^2$.
}
\label{chi_2d}
\end{figure}

\begin{figure}
\vspace{70mm}
\includegraphics{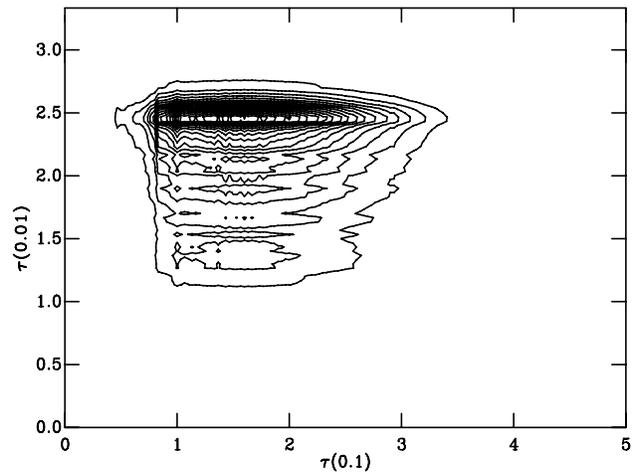}
\caption{
A two-dimensional differential probability distribution of $\tau$ for
the DAM97 isochrones and uncertainties of 0.01 mag (horizontal axis) and
0.1 mag (vertical axis).
The contours are evenly spaced starting at a probability of zero.
Note that the axes are in $\tau$ not $\tau^2$.
}
\label{tau_2d}
\end{figure}

The standard proof for the $\chi^2$ distribution for many datapoints is
a generalization of the proof for just two \citep[e.g.][]{saha}.
One considers a two-dimensional space, with $\chi_1$ (not
$\chi^2_1$) as one axis and $\chi_2$ as the other.
At each point in the space one evaluates the probability of obtaining
simultaneously values of $\chi_1$ between $\chi_1$ and $\chi_1 + d
\chi_1$ and of $\chi_2$ between $\chi_2$ and $\chi_2 + d
\chi_2$.
This probability is simply ${{dP_r}\over{d\chi_1}}{{dP_r}\over{d\chi_2}}$, or in 
more familiar terms of the differential probability distribution of 
$\chi^2$, $ 4 \chi_1 \chi_2 {{dP_r}\over{d\chi^2_1}}{{dP_r}\over{d\chi^2_2}}$.
This function is plotted in Figure \ref{chi_2d} using $P_r(\chi^2)$
for one degree of freedom.
The figure shows that the 
probability of obtaining any given value of $\chi^2 = \chi^2_1 + \chi^2_2$ is
independent of the value of either $\chi^2_1$ or $\chi^2_2$.
This allows the proof for $\chi^2$ to proceed to its conclusion that
the probability of obtaining any given value of $\chi^2$ is
proportional to $\chi_1 {{dP_r}\over{d\chi^2_1}}$ 
times the length of the arc at a radius $\chi^2$ (or more
generally the area of the N-dimensional surface).
The crucial point here is the interpretation that the probability of
obtaining a given $\chi^2$ is
simply the line integral of the probability along a line of constant
$\chi^2$.
For $\chi^2$ the integral can be performed analytically, because the
probability is the same along a line 
of constant $\chi^2$; this is not the case for $\tau^2$, and in this
case the integral must be evaluated numerically.

Figure \ref{tau_2d} shows the equivalent plot to Figure \ref{chi_2d},
but instead of  $\chi^2 = \chi^2_1 + \chi^2_2$ we have 
$\tau^2= \tau^2 (\sigma = 0.01) + \tau^2 (\sigma = 0.1)$, for the DAM97 
isochrones.
Before embarking on how to use this plot to determine the probability
of obtaining a given $\tau^2$ or greater, it is useful to understand
the differences between Figures \ref{chi_2d} and \ref{tau_2d}.
The most likely value of $\chi$ is zero, simply because the most
probable position for a datapoint to lie at its value before
perturbation by observational error.
For $\tau$ this is not the case.
At any given value of (say) $V$, there are a range of actual $V-I$
values it could have originated from.
Furthermore, the large area of the CMD covered by binaries (albeit at a low
probability), gives a very large chance that a star will yield a high
$\tau$.
This point can be emphasised in two ways.
First, collapsing the plot onto the y-axis gives the differential version of
the upper curve in Figure \ref{tau_integ}, with its emphasis on high
values of $\tau^2$.
Second, collapsing the curve onto the x-axis yields a distribution
more strongly skewed to low values, as one would expect because the
larger value of $\sigma$ causes the distribution to tend towards that
for $\chi^2$. 

The problem with Figure \ref{tau_2d}, from the point-of-view of
evaluating $P_r(\tau^2)$ is that ${d^2P_r}\over{d\tau_1 d\tau_2}$ along a line
of constant $\tau^2$ is not independent of either $P_r(\tau^2 (\sigma =
0.01))$ or $P_r(\tau^2 (\sigma = 0.1))$.
This precludes us using the analytical $\chi^2$-method to evaluate $P_r(\tau^2)$.
However, this does not stop us undertaking a numerical line
integration along fixed curves of $\tau^2$ to evaluate the probability
of exceeding that value of $\tau^2$.
The route we have followed to perform this numerical integration
relies on the fact that the arc length is proportional to the number
of pixels.
One can calculate a grid of the differential probability (i.e.  the
probability of obtaining a certain $\tau^2$, not of exceeding it),
akin to Figure \ref{tau_2d} by simply multiplying the two differential
distributions together.
A simple histogram of the number of pixels with a given value of 
$\tau^2$ is then ${dP_r}\over{d\tau^2}$.
Unfortunately, when one generalizes this to say, the 100 dimensions
needed for a 100 point dataset, the calculation becomes intractable in
reasonable computation times.
We therefore perform the calculation dimension by dimension.
We take the first two $\tau^2$ distributions, and multiply each point
in one distribution by every other point in the other.
We then bin the result into bins of $\tau^2 = \tau^2_1 + \tau^2_2$ to
produce a new, one dimensional distribution.
This can then be multiplied by the next dimension, and the process
repeated until all dimensions have been allowed for.
We then integrate this to change from a differential to a cumulative
distribution.

\subsubsection{$\tau_\nu^2$ and practicalities}
\label{tau_prac}

The method described thus far is very general, with little tailoring to
the specific problems of CMDs.
In calculating the expected distribution of $\tau^2$ however, we must
return to the subject of normalising-out the mass function, a
procedure first discussed in Section \ref{fit_and_par}.
If the model for Figure \ref{dam97_40_fig} included a mass function
with increasing numbers of stars at fainter magnitudes, we would
expect to see a much higher probability density in the bottom part of
the plot than in the top part.
Since we expect the majority of our datapoints to lie at faint
magnitudes, this is perfectly correct.
The best $\tau^2$ values
will be found by placing the majority of the points in the regions of
highest probability density; thus the mass function is a driving force 
in the fitting procedure.
Note, however, that the distribution of $\tau^2$ 
is different for bright and faint magnitudes, due
largely to the change in slope of the pre-main-sequence.
This means that the distribution of $\tau^2$ for a single datapoint is
different for different mass functions.
For the observational reasons explained in Section \ref{fit_and_par}
it is not desirable to introduce the mass function as a set of free
parameters, and so we have normalised-out the mass function in our
models by setting the integral of $\rho(m,c)$ over all colours at a
given magnitude to one.
This has the additional advantage that the distribution of $\tau^2$
reduces to that for $\chi^2$ for data with uncertainties in two
dimensions fitted to a straight line (Section \ref{special}).

Given that we are not interested in determining the mass function, just
the age and distance of clusters, how are we to calculate a
$\tau^2$ distribution in the case of a normalised-out mass function?
Our method is to calculate the distribution of $\tau^2$ for each data
point using only the region of the CMD within $\pm3\sigma_m$ of its
measured magnitude.
Thus the ensemble of individual $\tau^2$ probability distributions,
and hence the distribution for the fit as a whole reflects the actual
distribution of datapoints in $V$-band magnitude.
This has the incidental advantage of greatly speeding the calculation,
the limiting factor being smoothing the image by the uncertainty for
the datapoint, for which the run-time scales linearly with the
magnitude range used.

\begin{figure}
\vspace{70mm}
\includegraphics{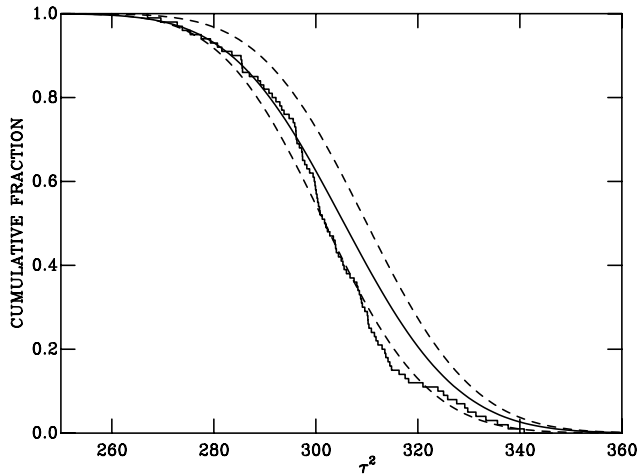}
\caption{
The distribution of $\tau^2$ obtained by fitting 100
simulated observations (histogram),
compared with the mean of the predictions for the distribution of
$\tau^2$ for each dataset (line). 
The dotted lines enclose the 50 percent of the $\tau^2$ predictions
with values at $P_r(\tau^2)=0.5$ closest to the mean, i.e.
the 25th to the 75th percentiles of the distribution.
The histogram is the same as that in Figure \ref{sim_tau}.
}
\label{mf_prob}
\end{figure}

This ``bootstrapping'' of the mass function leads to a fundamental
limit on how well we can predict $P_r(\tau^2)$.
Consider the hundred simulated observations of Section \ref{goodness}.
Because the datapoints are all slightly different, then for each
dataset we have a different prediction for the distribution
$P_r(\tau^2)$.
This situation is illustrated in Figure \ref{mf_prob} where the solid 
curve shows the mean of all 100 predictions.
To assess the range of predictions we sorted the distributions by the
value of $\tau^2$ at a probability of 0.5, and then plotted (as a
dotted lines) the distributions that enclosed in middle 50 percent of
these values, i.e. the 25th to the 75th percentiles of the distribution.
These cover a range of about 1 percent in $\tau^2$.
This indicates that the prediction for $P_r(\tau^2)$ from a single
observation (such as the solid line in Figure \ref{sim_tau})
is uncertain at the $\pm$1 percent level in $\tau^2$ which corresponds 
$\pm$0.1 in $P_r(\tau^2)$.

There is a final complication which adds a further uncertainty to the
absolute value of $\tau^2$.
The method outlined above only calculated the distribution 
$P_r(\tau^2)$ for matching the data directly to a model, with no free 
parameters.
In the $\chi^2$ case, for large values of the number of degrees of
freedom (i.e. the number of free parameters $n$ subtracted from the
number of datapoints $N$), $P_r(\chi^2)$ scales with the number of degrees
of freedom.
This implies that we need to multiply our $P_r(\tau^2)$ by $(N-n)/N$.
We have no formal proof for this, but the following numerical
experiment supports this view.
If one takes a simulated observation and compares it with the underlying
model one obtains a value for $\tau^2$.
If it is now compared with a grid of models with a range of distance
moduli and ages, the best fitting model will have a smaller value of
$\tau^2$.
Over many realisations we find the mean change is a factor of $(N-n)/N$.

In summary, therefore, we calculate the expected distribution of
$\tau^2$ by first considering one datapoint at a time, after the
fitting process is complete.
We smooth the best fitting distribution in colour-magnitude space
according to the uncertainties for that point, and then extract the
distribution of probability $P$ as a function $\tau^2 = - 2 {\rm ln}
P$.
We then multiply all the distributions together, using the method
described above, to find the expected distribution of $\tau^2$ for
our dataset.
Finally, we can divide our fitted value of $\tau^2$ (and the
values of $\tau^2$ in $P_r(\tau^2)$)
by the expected value of $\tau^2$ at $P_r(\tau^2)$=0.5.
In analogy with $\chi^2_\nu$ this gives us $\tau^2_\nu$, that has an
expected value of unity for a good fit.

\section{NGC2547 - A WORKED EXAMPLE}
\label{2547}

An important test of any algorithm is whether it will work with real,
as well as simulated data.
We have chosen as our test dataset the X-ray selected sample of
members of the young open cluster NGC2547, which we first fitted in
\cite{2002MNRAS.335..291N}. 
We have chosen this cluster as the dataset has already been fitted
by one of the authors using traditional ``by eye'' methods,
allowing us to make a direct comparison of the methods.

\subsection{Soft clipping}

The main practical problem which must be solved is that some of the
datapoints lie in regions of the CMD to which our model assigns
probabilities ($\rho(c,m)$) of zero. 
Of course, in principle no point on the CMD has zero probability, once
it is blurred by the uncertainties and becomes $P$.
Practically however, once one is a few $\sigma$ from the sequence
numerical rounding ensures that taking the logarithm of this
probability causes a numerical error.
The underlying philosophical issue here is that these datapoints are
probably not described by our model (they are background or foreground
contamination) and at some point these data should be removed from
the fitting process.
The classical way of dealing with such a situation is an $N \sigma$
clipping scheme, removing datapoints from the calculation of $\tau^2$
once they lie at very low probabilities ($N \sigma$ from the expected value).
Simple clipping would be a recipe for numerical instability, so
instead we use a soft clipping scheme.
To achieve this we simply add a small probability ($e^{-0.5\times20}$) to $P_i$
for each datapoint, the value we use amounting to a maximum $\tau^2$
of 20 for each datapoint.
We then search for the minimum in $\tau^2$ space using the full
dataset, but once the best fit has been found, we clip out all the
datapoints whose $\tau^2$ exceeds half the maximum $\tau^2$ set. 
It this subset for which we then
calculate the expected value of $\tau^2$ (see Section \ref{binaries}).

\subsection{Magnitude independent uncertainty}

In addition to the statistical uncertainty given for each datapoint,
\cite{2002MNRAS.335..291N} also point out that there is a
magnitude independent uncertainty for each datapoint, due to
uncertainties in the profile correction.
Essentially this is the uncertainty due to correcting the magnitude
measurements back to the large apertures required for standard stars.
This should be clearly distinguished from the error in
the transformation from the natural to the standard system, which has
the effect of shifting all the data points in the same direction.
We use a magnitude independent uncertainty of 0.01 mags for each
filter (thus 0.01 mags in $V$ and 0.014 in $V-I$) as a good
approximation to the magnitude independent uncertainty given by
\cite{2002MNRAS.335..291N}.
This is added in quadrature with the statistical uncertainties.
As we shall see below, this value is also justified by the fact that
we obtain a reasonable value for $P_r(\tau^2)$.
For datasets where this is not the case, one has the possibility of
adjusting the magnitude independent uncertainty until a $P_r(\tau^2)$
of approximately 50 percent is obtained.

\subsection{Extreme mass-ratio binaries}
\label{extreme}

A second issue is the absence from our models of extreme
mass-ratio binaries.
This was first pointed out in Section \ref{intuitive}, and is due to
the fact that the isochrones do not reach sufficiently low masses to allow
us to model the most extreme binaries.
In the simulations we have performed thus far this is not an issue as
both the simulated data and the models suffer from the same cut-off.
To simulate the case of real data we therefore created a new set of
models where the lowest mass stars available for the binaries were
0.25M$_\odot$ (compared with 0.017M$_\odot$ in the isochrones).  
We then fitted these models to simulated datasets with the underlying
parameters used in Section \ref{num_exp}, which therefore contained binaries
created using the full range of masses available in the isochrones.
The mean parameters from 30 simulated observations were
40.07$\pm0.16$Myr and a distance modulus of -0.002$\pm$0.001, where
the uncertainties are standard errors.
Thus the effect on the parameters of a low-mass cut-off for the
binaries is undetectable in our simulation, and certainly an
order-of-magnitude below our statistical uncertainties.


\subsection{Results}
\label{results}

\begin{figure*}
\vspace{95mm}
\includegraphics{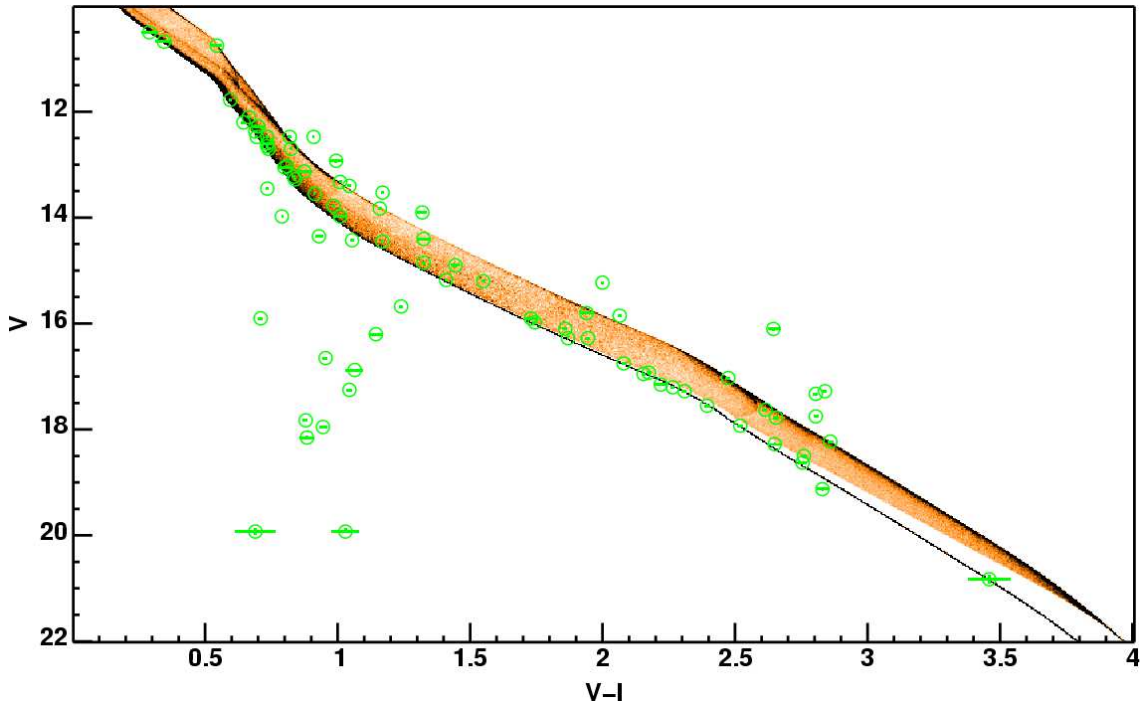}
\caption{
The X-ray selected members of NGC2547 (green circles with error bars)
and best fitting model (background image), for a binary fraction of 0.5.}
\label{best_model}
\end{figure*}

As in \cite{2002MNRAS.335..291N} we used the models of DAM97, and
extinctions of $E(V-I_C)$=0.077 and $A_V=0.186$.
We began by assuming 50 percent of the unresolved images are binaries, 50
percent single stars, and assumed (even when we changed the binary
fraction) that the masses for the secondary stars were uniformly distributed
between the mass of the primary and the lowest mass available in the models.
The best fit gives $P_r(\tau^2)$=0.41 and is shown in Figure
\ref{best_model}.
Considering this is a new technique, finding an acceptable value of
$\tau^2$ (equivalent to finding a reduced $\chi^2$ of approximately
one) is very encouraging both in terms of the verisimilitude of
the models, and of the accuracy of our observations.
The best fitting values and 67 percent confidence limits are
$38.5^{+3.5}_{-6.5}$Myrs and a true distance modulus of $7.79^{+0.11}_{-0.05}$.
Although the fit is good, close examination shows that between
$V$=13.5 and 15 the model seems to lie show systematically below the data.
First, it should be made clear that this effect is small (0.02 mags in
$V-I$).
Second, it might be thought that by decreasing the distance modulus
one could fit these points, and fit the lower pre-main-sequence by
choosing a slightly greater age.
In fact the models show that the region at $V=14$ is moving bluewards
with age faster than the lower part of the sequence, and the $\tau^2$
test has chosen a reasonable compromise.
The systematic residuals are, therefore, real differences between the
shape of the model isochrones and observed sequences.

\subsection{Changing the binary fraction}

\begin{figure}
\vspace{70mm}
\includegraphics{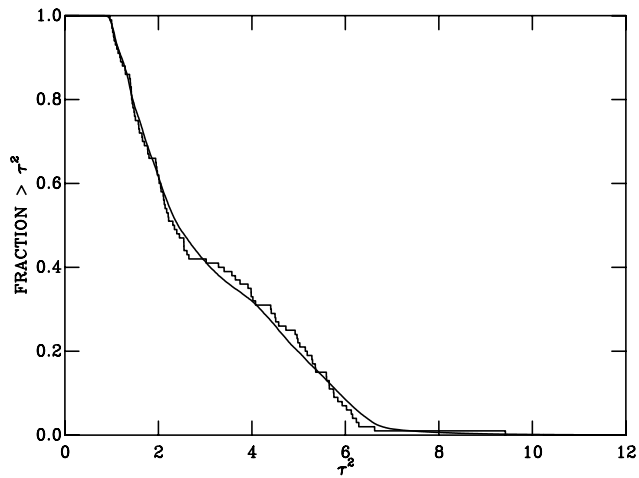}
\caption{  
The expected distribution of $\tau^2$ (curve) and that obtained from
the data (histogram) for the simulated dataset of Section \ref{num_exp}.
}
\label{just_for_rob}
\end{figure}

\begin{figure}
\vspace{70mm}
\includegraphics{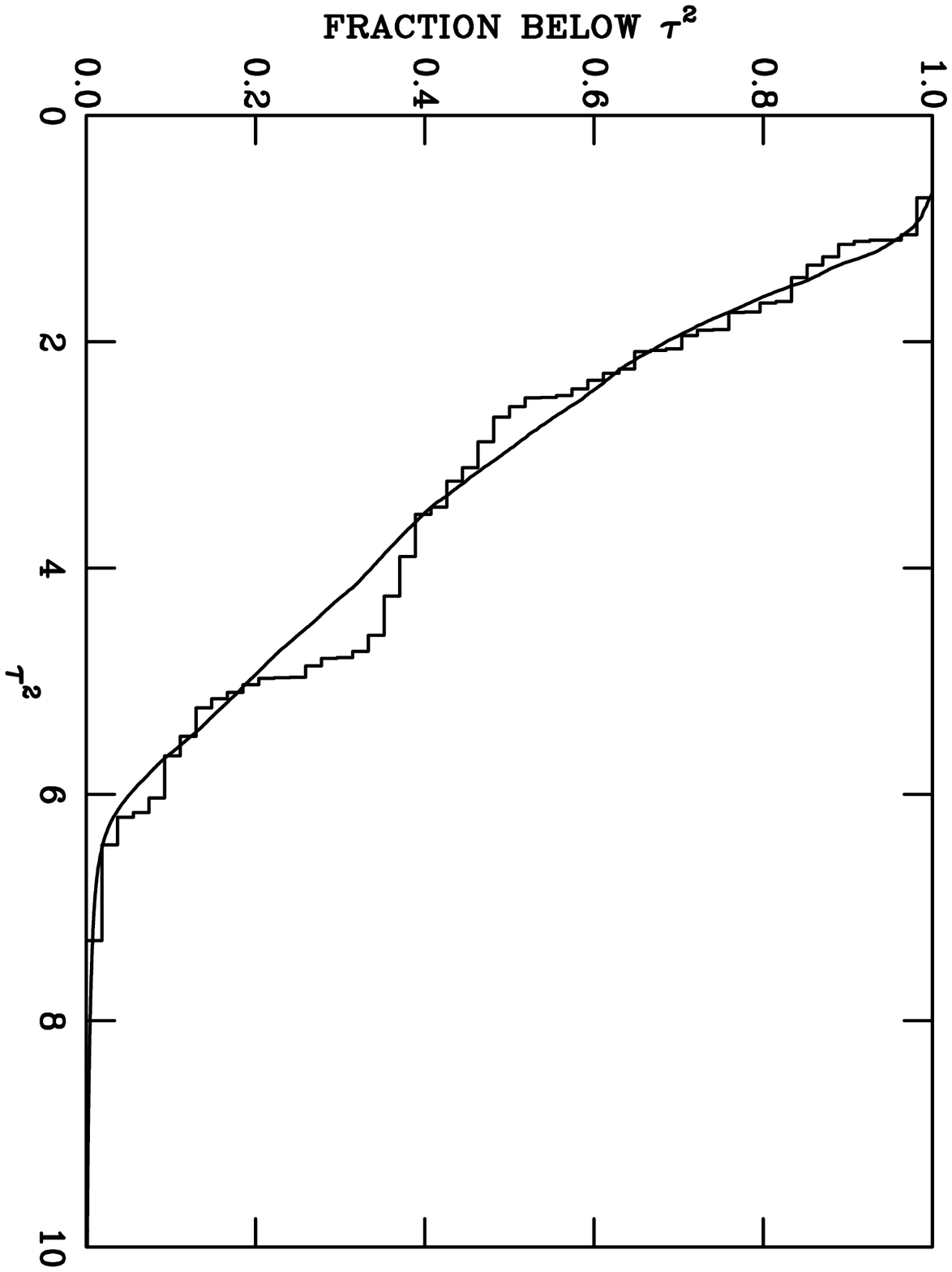}
\caption{  
The expected distribution of $\tau^2$ (curve) and that obtained from
the data (histogram) for NGC2547, assuming a binary fraction of 50
percent.
}
\label{tau_one_50}
\end{figure}

\begin{figure}
\vspace{70mm}
\includegraphics{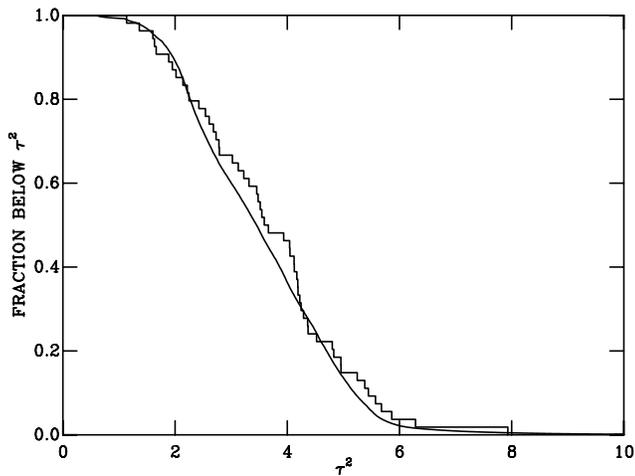}
\caption{
The expected distribution of $\tau^2$ (curve) and that obtained from
the data (histogram) for NGC2547, assuming a binary fraction of 80
percent.
}
\label{tau_one_80}
\end{figure}

Although Figure \ref{best_model} shows the that the single star
sequence is broadly correct, it is harder to assess the fit
to the binary stars.
The distribution of $\tau^2$ for the individual datapoints gives us a
useful insight into this.
In Section \ref{binaries} we described how we calculate the
probability distribution of $\tau^2$ for each datapoint before
multiplying them together to predict the overall value for $\tau^2$
for the fit.
Instead of multiplying them, the sum of the probability distributions 
gives us the expected distribution of $\tau^2$ for the datapoints in
the best fit.
Before carrying out a comparison with the NGC2547 data, we show in
Figure \ref{just_for_rob} the comparison between the actual
(histogram) and predicted (curve) distributions of the single-point 
$\tau^2$ values for the simulated cluster used in Section \ref{num_exp}.
This shows the prediction works very well.
In Figure \ref{tau_one_50} we show the same plot for NGC2547.
The real distribution differs from the model in the mid-ranges of
$\tau^2$, in particular there are more  points at 
$\tau^2\simeq4$ than the model predicts.
High values of $\tau^2$ correspond to low values of $P_i$.
The majority of the low values of $P_i$ will occur in the region
between the single-star and equal-mass-binary sequences, 
implying that we have underestimated the binary fraction.
To test this hypothesis, and to establish whether one must correctly
model the binary fraction to determine reliable ages and distances,
we re-fitted the data with a binary fraction of 80 percent.
The actual and predicted distributions of $\tau^2$ shown in Figure 
\ref{tau_one_80}.
Increasing the binary fraction has indeed increased the number of
high valued $\tau^2$ points, but in fact the model is now
systematically lower than the data.
Furthermore $P_r(\tau^2)$ is now only 0.14.
Clearly a binary fraction of 50 percent is a better fit to the data
than 80 percent.

There is a strong temptation at this point to attempt to model the
properties of the binaries, and indeed the ability to extract such 
information is one of our primary motivations for developing $\tau^2$.
However, it clearly lies outside the scope of this introductory paper
to do so.
Furthermore, in this case the dataset itself is unsuited to such
an experiment.
Aside from the question as to whether an X-ray selected sample is biased
towards binary stars, the reader should also note that some stars
appear above even the equal-mass-binary sequence.
Although some of these may be multiple systems with more than two
members (which we have ignored in our models), there are three times
more of them than we might expect from \cite{1991A&A...248..485D}.
For the majority of these objects, therefore, our result
implies that there is a non-photospheric contribution to their 
luminosity, which again would not be surprising for an X-ray selected
sample, or that we have significant contamination from foreground
dwarfs.
Either case would clearly preclude a photometric determination of
binary parameters.
Despite these cautions, it is interesting to note that we
obtain a credible value of $P_r(\tau^2)$ for a binary fraction which is
close to that determined by \cite{2002MNRAS.335..291N} (60 percent),
when they too assumed a flat mass ratio distribution.
Equally importantly, with a binary
fraction of 80 percent we obtained a distance modulus of 7.82 and an
age of 37.5Myr, which is not significantly different from the result
for a binary fraction of 50 percent.
The conclusion that the binary fraction has little effect on the
derived parameters is, in retrospect, unsurprising.
It means that the fit is being driven by the single-star sequence, and
not being dragged to brighter magnitudes by the binaries.

\subsection{Comparison with previous work}

\begin{figure}
\vspace{80mm}
\includegraphics{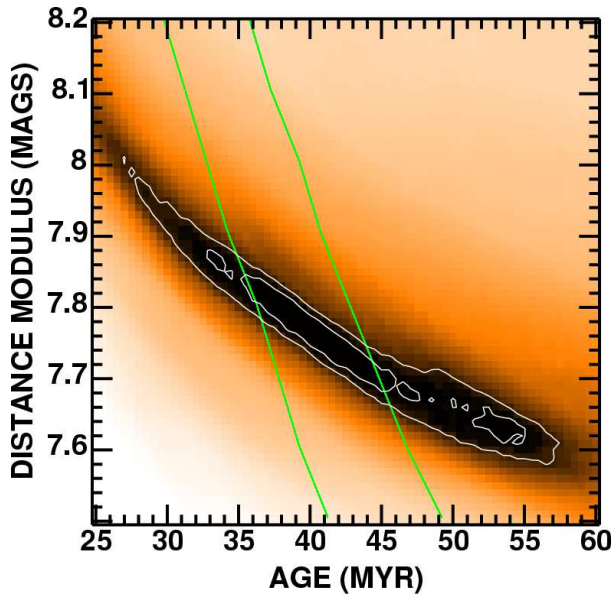}
\caption{
The $\tau^2$ grid for fitting the NGC2547 X-ray members to a
model with an 80 percent binary fraction.  
The minimum value of $\tau^2$ is 697.8 and 
the white contours are the 67 percent ($\tau^2$=704.6) and 95 percent
($\tau^2$=711.8) confidence limits.
The observed lithium depletion boundary requires the age and distance to lie
between the two green lines.
}
\label{grid}
\end{figure}

Although it is easiest to quote our result in terms of single
parameters and their uncertainties, the derived age and distance are
strongly correlated.
This is summarised in our $\tau^2$ space in Figure \ref{grid}.
Interestingly there is a second minimum (not as deep as the primary
one) at 53Myr and a distance modulus of 7.63 mags.
This is exactly the effect discussed in Section \ref{results}, and
when the fit is examined, we find that the data at bright magnitudes
lie systematically above the model.

The age/distance-modulus pair of 25$\pm$5Myr 8.05$\pm$0.10 
derived from the $V$/$V-I_C$ data in \cite{2002MNRAS.335..291N}
clearly lie in the $\tau^2$ valley of Figure \ref{grid}.
The position within that valley cannot be directly compared with
\cite{2002MNRAS.335..291N} as they used $B$/$B-V$ data to constrain the
distance.
Perhaps most remarkable is the excellent agreement with the lithium 
depletion age of \cite{2005MNRAS.358...13J}.
The age derived from the lithium depletion boundary depends on the
distance modulus.
Using the data of \cite{2005MNRAS.358...13J} and the DAM97 models we
derive an age of $37\pm3$Myr for our best-fit distance modulus of 7.8
mags.
However, we can also plot the constraint in Figure \ref{grid}, which
emphasises the concordance between the lithium and isochronal ages.
Our error bars in distance just fail to overlap at 1$\sigma$ with
those from HIPPARCOS ($8.18^{+0.29}_{-0.26}$) given by
\cite{1999A&A...345..471R}, but the distances are clearly not inconsistent.
Our conclusion is, therefore, that when used with real data $\tau^2$
fitting gives credible values and uncertainties.

\section{CONCLUSIONS}
\label{conclusions}

We have developed a maximum likelihood method for determining
parameters for an isochronal population which contains binaries, from
its colour-magnitude diagram.
We have used numerical simulation to demonstrate it is correct, and
used it on a practical example.
There is clearly scope for further development.
Most obviously one could search many more parameters than we have,
determining, for example, binary fraction and mass ratio distribution,
mass function, metallicity, or extinction.
Several of these could be allowed to vary simultaneously.

One could also use this as a search statistic, looking for populations
of a given age in large are surveys.
Here the absolute value of $\tau^2$ would measure how likely a given
``sequence'' is to have occurred by chance.
Furthermore, one could not only search an N-dimensional colour
magnitude space, but might also wish to use other parameters, such as
position on the sky modelled against a clustered distribution.
Finally there is also a range of other problems to which the
technique might be applied such as modelling mass segregation in the
mass-radius diagram \citep[e.g.][]{2003MNRAS.345.1205L}, and one could even
conceive of a replacement for the 1D Kolmogorov-Smirnov test where the
datapoints had associated uncertainties.

\section*{Acknowledgments}

We are grateful to Charles Williams for useful discussions, and Peter
Draper for help with the Gaia package, which we have used to present our
$\tau^2$ spaces and colour-magnitude diagrams.

\bibliographystyle{mn2e}
\bibliography{text}

\bsp

\label{lastpage}

\end{document}